\documentstyle[aps,pre,epsfig,multicol,psfig]{revtex}

\newcommand{\beq}    {\begin{equation}}
\newcommand{\eeq}    {\end{equation}}
\newcommand{\beqa}   {\begin{eqnarray}}
\newcommand{\eeqa}   {\end{eqnarray}}

\begin{document}

\draft

\widetext

\title{Measuring the Lyapunov exponent using quantum mechanics}

\author{F. M. Cucchietti,$^a$ C. H. Lewenkopf,$^b$ 
        E. R. Mucciolo,$^c$ H. M. Pastawski,$^a$ and R. O. Vallejos$^d$}

\address{$^a$Facultad de Matem\'{a}tica, Astronom\'{\i}a y
F\a'{\i}sica, Universidad Nacional de C\'{o}rdoba,\\ Ciudad
Universitaria, 5000 C\a'{o}rdoba, Argentina\\ $^b$Instituto de
F\'{\i}sica, Universidade do Estado do Rio de Janeiro, 20559-900 Rio
de Janeiro, Brazil \\ $^c$Departamento de F\'{\i}sica,
Pontif\a'{\i}cia Universidade Cat\'olica do Rio de Janeiro, CP 38071,
22452-970 Rio de Janeiro, Brazil \\ $^d$Centro Brasileiro de Pesquisas
F\'{\i}sicas, R. Xavier Sigaud 150, 22290-180 Rio de Janeiro, Brazil}

\date{\today}

\maketitle


\begin{abstract}
We study the time evolution of two wave packets prepared at the same
initial state, but evolving under slightly different
Hamiltonians. For chaotic systems, we determine the circumstances that
lead to an exponential decay with time of the wave packet overlap
function. We show that for sufficiently weak perturbations, the
exponential decay follows a Fermi golden rule, while by making the
difference between the two Hamiltonians larger, the characteristic
exponential decay time becomes the Lyapunov exponent of the
classical system. We illustrate our theoretical findings by
investigating numerically the overlap decay function of a
two-dimensional dynamical system.
\end{abstract}



\begin{multicols}{2}

\narrowtext

\section{Introduction}
\label{sec:introduction}

Over the last two decades the quest for quantum fingerprints of
classical chaotic behavior has been a key subject of investigation in
quantum chaos \cite{Haake00,Stockmann99}. As a result, signatures of the
classical underlying dynamics were identified in the spectra, wave
functions, and time evolution of a large set of quantum systems.
However, one of the simplest indications of classical chaos, namely
the Lyapunov exponent, remained unrelated to the quantum dynamics
\cite{Heller}.  A
clear advance in this direction has been made recently by Jalabert and
Pastawski \cite{Jalabert01}, who proposed that the classical Lyapunov
exponent is measured by the decay rate of an overlap between perturbed
and unperturbed quantum states evolving from the same initial
state. Their work triggered several numerical studies
\cite{Cucchietti01,Prozen01,Jacquod01,Cerruti01} whose results are not
always in line with the original predictions of
Ref. \onlinecite{Jalabert01}. The main goal of this paper is to
clarify the range of applicability of these predictions and to
understand under which conditions it is possible to extract a
classical Lyapunov exponent from the quantum evolution of a system.

The object of study is the comparison between the time evolution of a 
wave packet under a given system Hamiltonian $H_0$ and the corresponding
evolution for a different Hamiltonian $H=H_0+V$. Formally this can be
quantified by the overlap amplitude
\beq
\label{eq:defO(t)}
O(t) = \langle \psi | \exp(i Ht/\hbar) \exp(-iH_0 t/\hbar) |\psi \rangle
\eeq
where, for the initial state $|\psi\rangle \equiv |\psi(0)\rangle$,
it is chosen a the Gaussian wave packet
\beq
\label{eq:initialwavepacket}
\psi({\bf r}, t=0) =
\frac{1}{(\sqrt{\pi}\sigma)^{d/2}}
       \exp \!\left[ \frac{i}{\hbar} {\bf p}_0 \cdot ({\bf r} - {\bf r}_0)
       - \frac{ ({\bf r} - {\bf r}_0)^2}{2\sigma^2}\right],
\eeq
centered at ${\bf r}_0$ and with initial momentum ${\bf p}_0$. The
purpose of such parameterization is twofold: The initial momentum
${\bf p}_0$ sets the wave packet mean energy range at which we define
(classically) the Lyapunov exponent, whereas the choice of a Gaussian
wave packet (with finite width $\sigma$) makes the theoretical
considerations tractable within the semiclassical approximation.

The amplitude overlap in Eq.\ (\ref{eq:defO(t)}) can be interpreted in
the following two different, though formally equivalent, ways: (a) A
wave packet is prepared at the time $t=0$ and let to evolve under
$H_0$ till a time $t>0$. The resulting state is then propagated
backwards in time under the Hamiltonian $H$ till $t=0$. Under such
construction, $|O(t)|^2$ gives the return probability. This is the
picture described by Ref. \onlinecite{Jalabert01} which was inspired
by some recent NMR experiments \cite{Pastawski95,Usaj98}. These
experiments explore the scenario that, under certain circumstances, it
is possible to evolve backwards in time a complex quantum system. This
is in the spirit of the {\sl gedanken} experiment at the origin of the
Boltzmann-Loschmidt controversy \cite{Balian} and, for that reason,
we call $|O(t)|^2$ the Loschmidt echo. Since the system is not
isolated, $|O(t)|^2$ is expected to decay as $t$ increases. The
construction given by Eq.\ (\ref{eq:defO(t)}) can be regarded as a way
to capture the physical effect of coupling the system to a complex
time-dependent environment, and hence relate $|O(t)|^2$ to dephasing
\cite{dephasing,Zurek01}. (b) Alternatively, one can regard $O(t)$ as the
overlap amplitude of an initial state $|\psi\rangle$ propagated
forward in time under $H_0$, with the same initial state
$|\psi\rangle$ propagated with $H$. This interpretation is closely
related to the concept of fidelity
\cite{Peres84,Zoller97,Peres93,Jacquod01,Cerruti01}.

Let us now state the main finding of Ref. \onlinecite{Jalabert01}.
There it is was shown that, after a suitable averaging (which shall be
discussed in the foregoing section), the return probability or
fidelity can be separated into two contributions,
\beq
\label{eq:defM(t)}
M(t) \equiv \overline{|O(t)|^2} = M_1(t) + M_2(t),
\eeq
both described in the long-time limit as
\beq
\label{eq:decaimentos}
M_i(t) \propto \exp (- \alpha_i t).
\eeq
The decay rate $\alpha_1$ depends on the properties of the
perturbation $V = H - H_0$, while $\alpha_2$ is the classical Lyapunov
exponent associated to the dynamics of $H_0$, provided $V$ is
classically weak. Depending on $V$ and $\lambda$ the decay can be
dominated by either $M_1(t)$ or $M_2(t)$. In this study we show under
which conditions it is possible to extract $\lambda$ from the analysis
of the average fidelity $M(t)$.

The structure of this paper is as follows. In Section \ref{sec:model}
we describe the model we use to obtain $\lambda$ from the quantum time
evolution. Section \ref{sec:theory} presents the analysis of the
different decay processes that govern the fidelity $M(t)$. There we
show that $M_1(t)$ is nothing else than the Fermi golden rule. The
classical and quantum relevant scales to the problem are discussed. In
particular, we show under which circumstances is it possible to
observe the Lyapunov decay. The numerical results verifying a Lyapunov
decay for our dynamical system are presented in Section
\ref{sec:numerics}. We then conclude in section
\ref{sec:conclusions} by relating our findings to the recent papers
mentioned above.

\section{The Model}
\label{sec:model}

To investigate the dependence of the Loschmidt echo on the magnitude
of an external perturbation we use as the unperturbed system the
smooth ``billiard'' stadium introduced in
Ref. \onlinecite{Vallejos99,Ortiz00}. This model consists of a 
two-dimensional Hamiltonian $H_0={\bf p}^2/2m + U({\bf r})$ with the 
potential given by
\beqa
U({\bf r}) = U_0\! \times \left\{ \begin{array}{cc} \infty, & x < 0,
\\ (y/R)^{2\nu}, & 0 \le x < d, \\ \Big\{[(x-d)^2 + y^2]/R^2\Big\}^\nu, 
& x \ge d\,. \end{array} \right.
\eeqa
In addition, $U({\bf r}) = \infty$ whenever $y<0$.
The exponent $\nu$ sets the slope of the confining potential. For
$\nu=1$ the smooth stadium is separable and thus integrable. As the
value of $\nu$ is increased, the borders become steeper. In the limit
of $\nu \rightarrow \infty$, the stadium gains hard walls, becoming
the well-know Bunimovich billiard, one of the paradigms of classical
chaotic systems. (Actually, we consider a quarter of a stadium in 
order to avoid features related to parity symmetries. \cite{Haake00})
Thus, by varying $\nu$, we can tune the system dynamics from integrable 
to chaotic.

\begin{figure}
\setlength{\unitlength}{1mm}
\begin{picture}(60,120)(0,0)
\put(-5,-5){\epsfxsize=9cm\epsfbox{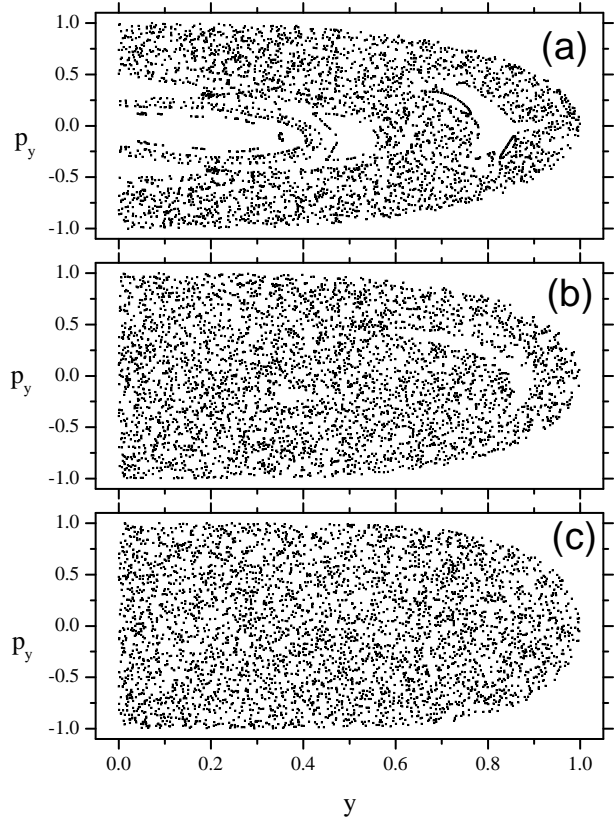}}
\end{picture}
\caption{Poincar\'e surface of section for the smooth stadium billiard
for $E=1$, $R=d=1$, and (a) $\nu=1.5$, (b) $\nu=2$, and (c) $\nu = 3$.}
\label{fig:Poincare1}
\end{figure}

In order to make the presentation more concise, we choose units such
that $U_0 = 1$ and $m=1/2$. Thus, for $R=d=1$ the equipotential
$U(x,y)=1$ corresponds to the border of the stadium with unit radius
and unit length. For any value of the energy $E$ the equipotential
$U(x,y)=E$ gives the classical turning points, defining the allowed
area ${\cal A} \equiv {\cal A}(E)$. This area is an important
parameter in the discussion of our numerical and analytical
results. Any exponent in the range $1<\nu\le 2$ already leads to a
mixed phase space, i.e., a situation with both regular and chaotic
motions present. In particular, for $\nu \ge 2$, $d=1$, and total
energy $E=1$ the classical dynamics is predominantly ergodic, although
small remnants of integrability still exist. These observations are
illustrated by the Poincar\'e surfaces of section displayed in
Fig. \ref{fig:Poincare1}.

\begin{figure}
\setlength{\unitlength}{1mm}
\begin{picture}(60,75)(0,0)
\put(4,-35){\epsfxsize=8cm\epsfbox{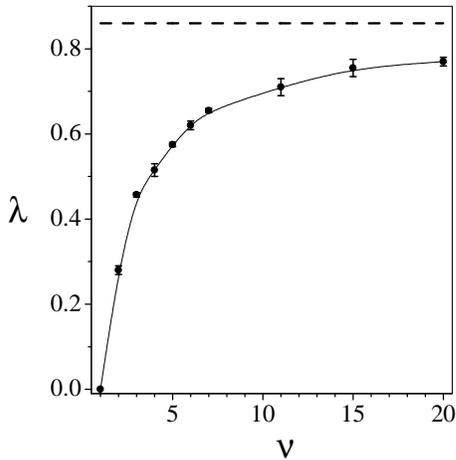}}
\end{picture}
\caption{The Lyapunov exponent of the smooth stadium for $E=1$ and
$R=d=1$ as a function of $\nu$. The circles are the results of our
computation, while the continuous line serves as a guide to the
eye. The dashed line corresponds to the billiard limit,
$\lambda_{\mbox{\scriptsize hard}} = 0.86$.}
\label{fig:lyapunov}
\end{figure}

The global Lyapunov exponent $\lambda$ for two-dimensional systems can
be easily computed by standard methods, such as that proposed by
Benettin {\it et al.} \cite{Benettin76}. The evolution of the
classical trajectories was carried out numerically using a symplectic
algorithm \cite{Yoshida90}.
We computed the Lyapunov exponent for several values of $\nu$. At
$E=1$, $\lambda$ varies smoothly as a function of $\nu$, as shown in
Fig.~\ref{fig:lyapunov}. As expected, as $\nu$ becomes very
large $\lambda$ approaches the value of the Lyapunov exponent for 
the Bunimovich stadium billiard, namely $\lambda_{\mbox{\scriptsize 
hard}} = 0.86$.

The work of Ref. \onlinecite{Jalabert01} used a Gaussian random
background potential as the perturbation that, once suddenly
switched on, mimics the effects of external sources of
irreversibility in the time evolution of a real system. Thus, static
disorder played the role of the external perturbation $V$. Our
strategy is essentially the same: we investigate $M(t)$ numerically
taking an ensemble average over different realizations of a disordered
potential $V({\bf r})$. For the later we choose a superposition of
${\cal N}_i$ independent Gaussian impurities, as in
Refs. \onlinecite{Jalabert01,Richter96a}:
\beq
V({\bf r}) = \sum_{j=1}^{{\cal N}_i} \frac{u_j}{2\pi\xi^2}
\exp\left[-\frac{|{\bf r} - {\bf R}_j|^2}{2\xi^2}\right].
\eeq
The vector ${\bf R}_j$ denotes the position of the $j$th impurity. All
impurities are uniformly distributed over an area ${\cal{A}}$ of the
two-dimensional plane where the stadium resides, with concentration
$n_i = {\cal N}_i/{\cal{A}}$. The strengths $u_j$ are Gaussian
distributed and uncorrelated, i.e., $\overline{u_j u_{j^\prime}} = u^2
\delta_{j j^\prime}$, with $\overline{u_j } = 0$.
The impurity potential defined above is statistically characterized by
the correlation function
\beqa
\label{eq:UUcorrelation}
C(|{\bf r} - {\bf r}^\prime|) & \equiv & \overline{
V({\bf r}) V({\bf r}^\prime)} \nonumber
\\ & = & \frac{u^2 n_i}{4\pi \xi^2} \exp\! \left[ - \frac{|{\bf r} -
{\bf r}^\prime|^2} {4 \xi^2} \right].
\eeqa
Notice that impurity averaging yields $\overline{V({\bf r})} = 0$.

\section{Theoretical background}
\label{sec:theory}

This section is devoted to the analysis of the time dependence of the
fidelity $M(t)$, explaining the origin of its different decay laws. We
discuss the relation between the decay regimes associated to $M(t)$
and the different time and perturbation strength scales of the
system. These considerations solve the recent controversy between
Lyapunov versus Fermi golden rule decay \cite{Jacquod01,Cerruti01}.

Let us start giving a more precise definition to $M_i(t)$ appearing in
Eq.\ (\ref{eq:defM(t)}), namely,
\beq
M_1(t) \equiv \left|\overline{O(t)}\right|^2 \quad \mbox{and} \quad
M_2(t) \equiv \overline{|O(t)|^2} -  \left|\overline{O(t)}\right|^2 \;.
\eeq
As it was already shown semiclassically \cite{Jalabert01}, both
$M_1(t)$ and $M_2(t)$ exhibit an exponential decay in time, but
different in nature. We show in the sequel that the prevailing decay
law is determined by the perturbation strength, as well as the time
range under consideration.

\subsection{The semiclassical approximation scheme}
\label{sec:semiclassicaltheory}

The best way to identify in $O(t)$ manifestations of the classical
underlying dynamics is to use a semiclassical approximation. This is
the essence of Ref. \onlinecite{Jalabert01}, which presents a complete
calculation scheme for $O(t)$ in the case of a chaotic $H_0$ and a
``weak" perturbation $V$. The starting point is the Gutzwiller
semiclassical propagator, casted in terms of a sum over all classical
trajectories $s$ going from ${\bf r}^\prime$ to ${\bf r}$ in the time
interval $t$:
\beq
\label{eq:semiclassicalK}
K^{V}({\bf r}, {\bf r}^\prime; t) = \!
  \sum_{s({\bf r},{\bf r}^\prime; t)} \frac{C_s^{1/2}}{2\pi i \hbar}
  \exp \!\left[ \frac{i}{\hbar} S_s^{V}({\bf r}, {\bf r}^\prime;t)
    - \frac {i \pi}{2}\mu_s \right],
\eeq
where $S^{V}$ denotes the action 
given by the integral of the Lagrangian, $S_s^{V}({\bf r}, {\bf
r}^\prime; t) = \int_0^t dt^\prime L(V)$. The superscript $V$ stands
for the perturbation potential, its absence indicates $V=0$. The
Maslov index corresponding to the trajectory $s$ is given by $\mu_s$
and $C_s = |\det(\partial^2 S_s/\partial r_i^\prime \partial r_j)|$
accounts for the conservation of classical probability in going from
the initial to the final position components $i$ and $j$,
respectively. To proceed analytically, it is necessary to restrict the
calculations to a situation where it is possible to neglect the
influence of the perturbation $V$ in the coefficients $C_s$.
In general,
the propagator $K^{V}({\bf r}, {\bf r}^\prime; t)$ describes the
quantum evolution problem with great accuracy up to very long times,
though shorter than the Heisenberg time \cite{Tomsovic91}. Since the
features we are interested in are manifest in a short time scale,
the semiclassical propagator is an adequate approximation.

We use $K^V({\bf r}, {\bf r}^\prime; t)$ to propagate the wave packet
$\psi({\bf r}^\prime, t=0)$ given by Eq.(\ref{eq:initialwavepacket})
at $t=0$ up to an arbitrary time $t$. After a simple integration, one
obtains
\beq 
\label{eq:psiV}
\psi^{V}({\bf r}, t) = \sqrt{4 \pi \sigma^2}\!\!\sum_{s({\bf r},
{\bf r}_0; t)}\!\!  K_s^{V} ({\bf r}, {\bf r}_0; t) \exp\! \left[ -
\frac{\sigma^2}{2\hbar^2} ({\bf p}_s - {\bf p}_0)^2 \right], 
\eeq
where ${\bf p}_s$ is defined by $\partial S/\partial{\bf
r}^\prime|_{{\bf r}^\prime= {\bf r}_0} = - {\bf p}_s$. Equation
(\ref{eq:psiV}) is obtained under the assumption $\xi \gg \sigma \gg
k^{-1}$, constraining the initial wave packet to be spatially
concentrated over a region smaller in diameter than the correlation
length of the fluctuations in $V({\bf r})$.

We can now calculate the overlap $O(t)$ as defined by
Eq. (\ref{eq:defO(t)}) by writing an analytical semiclassical
expression for $\langle \psi^V(t)|\psi(t)\rangle$. For times shorter
than the Heisenberg time, this is possible through the diagonal
approximation \cite{Jalabert01}. This approximation is standard
\cite{Ozorio98} and neglects contributions from pairs of trajectories
which are different, namely, $s \neq s^\prime$.  The resulting
expression reads
\beqa
\label{eq:Osc}
O(t) & = & \frac{\sigma}{\pi \hbar^2} \int \!d{\bf r} \, \sum_{s({\bf
r}, {\bf r}_0; t)} C_s \exp \!\left( \frac{i}{\hbar} \Delta S_s
\right) \nonumber \\ & & \times \exp\!\left[ - \frac{\sigma^2}{\hbar}
(\overline{\bf p}_s - {\bf p}_0)^2\right], 
\eeqa
where the action difference $\Delta S_s$ is just 
\beq
\label{eq:DeltaS}
\Delta S_s = - \int_0^t dt^\prime \, V[{\bf q}_s(t^\prime)].
\eeq 
Notice that phase difference accumulated along a trajectory $s$ is
solely due to the perturbation potential $V$.
 
At this level, the fidelity $M(t)$ is trivially written by taking the
modulus squared of $O(t)$, which implies in summing over pairs of
trajectories $s$ and $s^\prime$ taking into account the interference
between phases, $(\Delta S_s - \Delta S_{s^\prime})/\hbar$. It is easy
to check that $V=0$ leads to $M(t)=1$, as expected
\cite{Jalabert01}. The double sum we refer to can be split in two
kinds of terms: (a) the diagonal ones, when the trajectories $s$ and
$s^\prime$ remain close to each other, and (b) the off-diagonal terms,
corresponding to an unrelated pair of trajectories $s$ and
$s^\prime$. In Ref. \onlinecite{Jalabert01} it was shown that after
disorder averaging the {\sl diagonal} contribution renders
\beq
\label{eq:M2sc}
M_2(t) \propto \frac{1}{t} \exp (- \lambda t), 
\eeq
where $\lambda$ is the classical Lyapunov coefficient. In the long time
limit, $t \gg 1/\lambda$, the exponential decay dominates and $M_2(t)$ 
reduces to Eq. (\ref{eq:decaimentos}). It is not
within our scope to give details of this derivation, but it is worth
mentioning that, after impurity averaging
\cite{Richter96a,Richter96b}, the calculations leading to
Eq. (\ref{eq:M2sc}) rely solely on generic assumptions about the
classical dynamics of $H_0$.

The contribution to the fidelity coming from off-diagonal terms,
$M_1(t)$, can be computed using the impurity average technique of
Refs. \onlinecite{Richter96a,Richter96b}. It amounts to computing the
variance of the phase appearing in Eq. (\ref{eq:Osc}). Assuming that
$\Delta S_s$ are Gaussian distributed, which is reasonable for
trajectories longer than $\xi$, one readily writes
\beq
\label{eq:Gaussianactions}
\overline{\exp\!\left(\frac{i}{\hbar}\Delta S_s \right)} =
\exp \left( -\frac{1}{2 \hbar^2} \overline{\Delta S_s^2}\right)\,,
\eeq
where, by recalling Eq.\ (\ref{eq:DeltaS}), the impurity average
$\overline{\Delta S_s^2}$ is written as
\beq 
\overline{\Delta S_s^2}= \int_0^t \! dt^\prime \!  \int_0^t \!
dt^{\prime\prime} \,C\large[r(t^\prime, t^{\prime\prime})\large].
\eeq
The distance in the impurity autocorrelation function $C$ is
$r(t^\prime, t^{\prime\prime}) = |{\bf q}_s(t^\prime) - {\bf
q}_s(t^{\prime\prime})|$. It is useful to change integration variables
to the center-of-mass $(q + q^\prime)/2$ and difference $q - q^\prime$
coordinates, with $q = v_0 t$ and $q^\prime = v_0 t^\prime$. For $\xi
\ll \sqrt{\cal A}$, it is a good approximation to extend the integral
over the coordinate difference to infinity.  We can make further
analytical progress if we specialize the discussion to hard-wall
billiard systems, which are good approximations to our model
Hamiltonian, particularly as $\nu$ is increased. In this case, the
integral over $(q + q^\prime)/2$ yields $L = v_0 t$. As a result, one
obtains \cite{Jalabert01}
\beq
\label{eq:semiclassicalFGRdecay}
M_1^{\rm sc}(t) \propto \exp(-\alpha_1 t), \quad \mbox{with} \quad
\alpha_1 = \frac{u^2 n_i}{2 \sqrt{\pi} \hbar^2 v_0 \xi} \,.
\eeq
Notice that the Gaussian ansatz for $\Delta S_s$ is not justified for
very short times in the range of $\xi/v_0$, which, in our case is of the
same order as $\tau \equiv \sqrt{\cal A}/v_0$. Thus, we are unable to
make predictions about $M_1(t < \tau)$, and, consequently, about the
constant factor multiplying $\exp(-\alpha_1 t)$ in Eq.
(\ref{eq:semiclassicalFGRdecay}). The exponential decay can also be
characterized by the typical length at which the quantum phase is not
modified by the presence of impurities,
\beq 
\label{eq:ell}
\ell = \frac{2 \sqrt{\pi} \hbar^2 v_0^2 \xi}{u^2 n_i} =
\frac{v_0}{\alpha_1}.
\eeq
This quantity is known as the elastic mean free path. 
Equation (\ref{eq:ell}) corrects a minor mistake in $\ell$ given by 
Refs.~\onlinecite{Richter96b,Jalabert01}, 
namely a missing factor of 1/2. \cite{mistake}
In the sequel we show the relation between this semiclassical result and 
the stochastic theory.

\subsection{The random matrix approach}
\label{sec:RMtheory}

The computation of $M_1(t)$ by the statistical approach is a standard
random-matrix result (see, for instance, Ref.\onlinecite{Agassi75} or
Appendix B of Ref.\onlinecite{Lutz99}). A somewhat similar calculation
was also recently carried out by Mello and collaborators
\cite{Gruver97}. Notwithstanding, it is instructive to describe how
this is done. The connection to the random matrix theory is made by
the Bohigas' conjecture \cite{Bohigas84} and the fact that the
classical dynamics of $H_0$ is chaotic. Consequently, the matrix
elements
\beq 
\label{eq:V_nn'}
V_{nn^\prime} = \langle n | V({\bf r}) | n^\prime
\rangle 
\eeq 
with respect to the eigenstates of $H_0$ are Gaussian distributed,
regardless the form of $V({\bf r})$. With this in mind, we can
calculate the averaged propagator
\beq 
K(t) = e^{-iHt/\hbar}\theta(t) \;. 
\eeq 
This task is usually carried out in the energy representation by
introducing the Green function operator
\beq
G(E) = \frac{1}{E + i \eta - H} \;, \quad \mbox{with}
\quad \eta \rightarrow 0^+ \;.
\eeq
The formal expansion of $G$ in powers of $V$ and the rules for
averaging over products of Gaussian distributed matrix elements give
\beq 
\overline{G} = G_0 \frac{1}{1 - \overline{VG_0V}G_0}\,,
\eeq 
where $G_0 = (E + i \eta - H_0)^{-1}$. The matrix representation
of $\overline{G}$ is particularly simple. In the eigenbasis of $H_0$
it becomes
\beq \overline{G}_{nn^\prime}(E) = 
\frac{\delta_{nn^\prime}} {E + i \eta - E_n - \Sigma_n (E)},
\eeq
where $E_n$ is the $n$-th eigenvalue of $H_0$ and 
\beqa 
\Sigma_n (E) = \sum_{n^\prime} \overline {V^2_{nn^\prime}}
(G_0)_{n^\prime} \equiv \Delta_n(E) - \frac{i}{2} \Gamma_n(E),
\eeqa
with
\beqa 
\Delta_n(E) & = & \mbox{PV} \, \sum_n
\frac{\overline{V^2_{nn^\prime}}} {E - E_n} \nonumber \\ \Gamma_n(E)
& = & 2\pi \sum_n \overline{V^2_{nn^\prime}} \delta(E - E_n)\,.
\eeqa
Here PV stands for principal value. The real part $\Delta_N(E)$
only causes a small shift to the eigenenergy $E_n$ and will thus be
neglected. Whenever the average matrix elements
$\overline{V^2_{nn^\prime}}$ show a smooth dependence on the indices
$n$, it is customary to replace $\Gamma_n$ by its average value,
\beq
\label{eq:GammaFGR}
\Gamma = 2\pi \overline{V^2}/\Delta,
\eeq
where $\Delta$ is the mean level spacing of the unperturbed
spectrum. In most practical cases, $\Gamma$ and $\Delta$ can be
viewed as local energy averaged quantities. Hence, the average
propagator in the time representation becomes
\beq
\label{eq:avKRMT}
\overline{K}_{nn^\prime}(t) = \delta_{nn^\prime}
\exp\!\left(-i \frac{E_n t}{\hbar} - \frac{\Gamma t}{2\hbar}
\right) \theta(t) \,.
\eeq
It worth stressing that $\Gamma$ arises from a
nonperturbative scheme; nonetheless, it is usually associated to the
Fermi golden rule due to its structure.

The average propagator obtained in Eq.\ (\ref{eq:avKRMT}) is easily
related to $M_1(t)$ by calculating $\langle\psi|\overline{K}|\psi
\rangle$. This step gives us also a more precise meaning to the smooth
energy dependence of $\Gamma(E)$: In our construction the latter has
to change little in the energy window corresponding to the energy 
uncertainty of $\psi({\bf r}, t)$, which is determined by $\sigma$. 
Thus, the RMT final expression for $M_1(t)$ is
\beq
\label{eq:RMTFGRdecay}
M_1^{\rm RMT}(t) = \exp(-\Gamma t/\hbar),
\eeq
with $\Gamma$ given by Eq. (\ref{eq:GammaFGR}). Equation
(\ref{eq:RMTFGRdecay}) does not hold for very short times, since we
neglected the smooth energy variations of $\Gamma_n$ and
$\Delta_n$. It is beyond the scope of RMT to remedy this situation,
since for that purpose nonuniversal features of the model have to be
accounted for.

Despite sharing the same formal structure, it remains to be shown that
both semiclassical and random model theory are strictly
equivalent. This is what we do next by deriving an expression for the
Fermi golden rule in terms of the classical quantities used in
Eq. (\ref{eq:semiclassicalFGRdecay}).

For chaotic systems, we can calculate the average off-diagonal
perturbation matrix elements using the universal autocorrelation
function of eigenstates first conjectured by Berry \cite{Berry77}. For
two-dimensional billiards this function reads
\beq
\label{eq:psipsiBerry}
\left\langle \psi_n({\bf r}_1) \psi_n({\bf r}_2)
\right\rangle = \frac{1}{\cal{A}} J_0 (k_n|{\bf r}_1 -
{\bf r}_2|),
\eeq
where $J_0(x)$ is the Bessel function of zero order, $k_n$ is the wave
number associated to the $n$th eigenstate of $H_0$, and ${\cal A}$ is
the billiard area. Here $\langle \cdots \rangle$ can be regarded as
the average $\psi_n({\bf r}_1) \psi_n({\bf r}_2)$ obtained by sweeping
${\bf R} = ({\bf r}_2 + {\bf r}_1)/2$ over a region containing several
de Broglie wave lengths. Equivalently, one could also average over a
large number of levels, provided that $k_n$ does not change much on
that interval. For a rigorous discussion on the validity of
Eq. (\ref{eq:psipsiBerry}) and the different averaging procedures, see
Ref. \onlinecite{Toscano01}.

Recalling Eq.\ (\ref{eq:V_nn'}), we can write the off-diagonal squared 
matrix elements averaged over the impurity realizations as
\beqa
\label{eq:averagebruto}
\overline{V_{n n^\prime}^2} = \int \!d^2r_1 \int\! d^2r_2\,&&
\psi_n         ({\bf r}_1)\psi_n         ({\bf r}_2)
\psi_{n^\prime}({\bf r}_1)\psi_{n^\prime}({\bf r}_2) \nonumber\\
&& \times \overline{V({\bf r}_1) V({\bf r}_2)}\;.
\eeqa
By changing variables to ${\bf R} = ({\bf r}_2 + {\bf r}_1)/2$ and
${\bf r} = {\bf r}_2 - {\bf r}_1$ and with the help of Berry's
conjecture, it is straightforward to reduce the integral in Eq.\
(\ref{eq:averagebruto}) to
\beq
\overline{V_{n n^\prime}^2} = \frac{1}{{\cal A}} \int \! d^2r \,
J_0(k_n r)J_0(k_{n^\prime} r)\,C(r),
\eeq
The correlation function $C$ is given by
Eq. (\ref{eq:UUcorrelation}). For a sufficiently large billiard, $\xi
\ll {\cal{A}}^{1/2}$, we obtain \cite{Gradshteyn94}
\beq
\overline{V_{n n^\prime}^2} \approx \frac{n_i u^2}{{\cal{A}}} e^{
-(k_n^2+ k_{n^\prime}^2)\xi^2} I_0(2 k_n k_{n^\prime} \xi^2) \;,
\eeq
where $I_0(x)$ is the modified Bessel function of the first kind. For
high-energy eigenstates, such that $k_n\xi \gg 1$, and for states
within an energy window corresponding to $\sigma$ ($k_n \approx
k_{n^\prime}$), the above expression is further simplified to
\beq
\overline{V_{n n^\prime}^2} \approx \frac{n_i u^2}{ {\cal{A}}}
\frac{1}{2 \sqrt{\pi}\, k_n \xi}\,.
\eeq
We can now insert $\overline{V_{n n^\prime}^2}$ into the left-hand
side of Eq. (\ref{eq:GammaFGR}). Recalling that the mean level spacing
for a two-dimensional billiard is $\Delta = 2\pi \hbar^2/({\cal{A}}
m)$ and using $\hbar k = m v_0$, we obtain
\beq 
\frac{\Gamma
}{\hbar} = \frac{u^2 n_i}{2
\sqrt{\pi} \hbar^2 v_0 \xi} \,.
\eeq
This is exactly the same decay rate of Eq. (\ref{eq:semiclassicalFGRdecay}). 
It also agrees with the quantum diagrammatic perturbation theory for the
bulk in the disordered model. \cite{Richter96b}

\subsection{Fermi golden rule and Lyapunov decay}
\label{sec:FermiLyapunovdecay}

By employing the semiclassical approach we were able to address in
detail two very distinct regimes of $M(t)$. Such approximation is the
most appropriate tool to study $M(t)$ provided two conditions are met:
$V$ is (a) classically weak, in the sense that classical perturbation
theory applies, and (b) quantum mechanically strong, meaning that one
can treat the actions in Eq.\ (\ref{eq:Gaussianactions}) as Gaussian
variables. In such cases, for $t \gg \lambda^{-1}$, it was found
that: (a) $\log M_2(t) \propto - \lambda t$ independently of the
strength of the perturbation and (b) $\log M_1(t) \propto - \Gamma t$,
where $\Gamma \propto u^2$. As one varies the perturbation strength
$u$, $M(t)$ is dominated by the smallest of $\lambda$ and $\Gamma$.
In other words, within the semiclassical regime, for small values of
$u$ the Fermi Golden Rule applies. The dependence of $M(t)$ crosses
over to $\log M(t) \propto - \lambda t$ when $\Gamma > \lambda$.
Equations (\ref{eq:M2sc}) and (\ref{eq:semiclassicalFGRdecay}) predict
for which value of $u$ this transition occurs.

It remains to be discussed what happens to $M(t)$ when $u$ does not
meet neither (a) nor (b), namely either $u$ is below the Fermi Golden
Rule regime or $u$ is in the very opposite limit of strong
perturbations, above the Lyapunov regime.

Let us first discuss the limit of ``extremely" weak perturbations,
where $V$ neither significantly mixed the states of $H_0$, nor causes
level crossings \cite{Szafer93,Simons93}. Here, $M(t)$ can be obtained
by standard quantum perturbation theory. This limit was studied a
long time ago by Peres \cite{Peres84}, who found a Gaussian decay,
namely $\log M(t) \propto - (ut)^2$. It turns out, as illustrated
by our numerical study, that this limit is very hard to observe, since
for very short times $M(t)$ decays as $t^2$ in all cases.

At the opposite end there is the case of ``strong" perturbations, for
which classical perturbation theory breaks down. As $u$ is augmented
the Lyapunov exponents of $H_0$ and $H$ become increasingly different. 
Lacking a theoretical understanding for this regime, we can only speculate 
that $M(t)$ decays faster than in the Lyapunov regime.
Here $M(t)$ will strongly depend on the specific details of $V({\bf r})$.

\begin{figure}
\setlength{\unitlength}{1mm}
\begin{picture}(60,50)(0,0)
\put(5,5){\epsfxsize=7cm\epsfbox{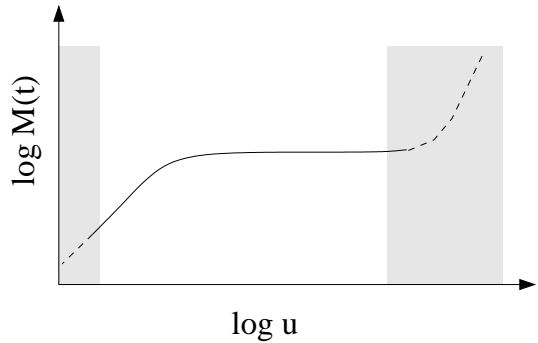}}
\end{picture}
\caption{Sketch of the expected behavior for $\log M(t)$ as a
function of the perturbation strength $u$ for a fixed value of
$t$. The shaded fields indicate the regimes of $u$ where the
semiclassical approach fails.}
\label{fig:phasediagram}
\end{figure}

Figure \ref{fig:phasediagram} summarizes the principal predictions of
this section. The main feature of this diagram is the plateau in
$-\log M(t)$ versus $u$, characterizing the Lyapunov regime. For a
given specific system we can predict where the plateau starts at low
values of $u$. To use a quantum system to measure the Lyapunov
exponent, it is crucial to know where it ends, and classical
perturbation theory breaks down. For that purpose, numerical
simulations were performed for the smooth stadium by varying its
classical Lyapunov exponent $\lambda$ and the perturbation strength
$u$. The results are presented in the next section.

\section{Numerical results}
\label{sec:numerics}

In this section we present a numerical study of $M(t)$ for the smooth
stadium model with Gaussian impurity disorder introduced in Section
\ref{sec:model}. Before showing the results, however, we describe some
technical details about the numerical method employed in the
simulations.

The quantum evolution of wave packets, as defined by
Eq. (\ref{eq:initialwavepacket}), was carried out through the
fourth-order Trotter-Suzuki algorithm \cite{TrotterSuzuki}. It is
worth noticing that a more straightforward approach, based on a matrix
representation of the evolution operator in terms of the eigenvectors
of $H_0$ would be far less efficient.

The method does not require spatial discretization of the
system. However, the basis has to be such that the Hamiltonian
matrix elements needs to involve only short term interactions. We
thus found useful to work on a lattice and to represent the kinetic
energy with a nearest-neighbor hopping term. Within the energy range
we explored, we found that a two-dimensional lattice of area $2.1 R
\times 1.1 R$ provided very accurate results when we employed $N=180$
sites per unit distance $R$ (with the intersite distance given by
$a=R/N$), corresponding to a total number of $378 \times 198$ lattice
sites.

The range of parameter values explored in our simulations was
limited by computational cost. Moreover, our choice of parameters
was guided by the constraints imposed by the semiclassical
calculations of Sec. \ref{sec:semiclassicaltheory}. First, in order
to include a large number of randomly located impurities, their
correlation width $\xi$ had to be taken much smaller than $R$. Second,
the semiclassical regime where Eq. (\ref{eq:decaimentos}) applies
requires $\xi$ to be larger than the wave packet width $\sigma$,
which, in turn, has to be much larger than the particle wavelength
$\lambda_F$. Other constraints arise from finite size effects. For
instance, the large-time saturation value of the Loschmidt echo, $M(t
\rightarrow \infty )$, depends on the ratio $\sigma/N$. Thus, for a
fixed $N$, it is necessary to make $\sigma$ as small as possible in
order to guarantee a small value for $M(t \rightarrow \infty)$. In
addition, let us recall that the energy spectrum of the (open
boundaries) discretized system is given by
\beq E_{\bf k} = \frac{2\hbar^{2}} {m a^{2}} -
\frac{\hbar^{2}}{m a^{2}} \left[ \cos(k_x a) + \cos(k_y a) \right] \,.
\eeq 
Therefore, we can only accurately recover the dispersion relation of
the free particle, $E_{\bf k} = \hbar^{2}k^{2}/2m$, when $k a \ll 1$.
All these constraints are summarized by the inequalities
\begin{equation}
\label{constraints}
a \ll \lambda_F \ll \sigma < \xi \ll R.
\end{equation}

The compromise between a good accuracy and a feasible simulation time
led us to set $\xi = 0.25R$, $\sigma = 0.18R$, $\lambda_F = 0.07R$,
and $N = 180$. This choice, combined with the values of the classical
model parameters, $m = 1/2$ and $E = 1$, gave raise to units such that
$\hbar=0.011 R$. Thus, the inequalities of Eq. (\ref{constraints})
were approximately observed in our simulations. For the quantum
evolution, a time step $\delta t = 2 m a^{2}/10 \hbar = 2.8 \times
10^{-4} E/\hbar$ proved to be sufficiently small.

It is important to make a few remarks about the averaging
procedure. In the simulations, besides averaging over impurity
configurations, we also found important to average over initial
positions ${\bf r}_0$ and directions ${\bf p}_0$. The main reason is
that numerical simulations of billiards deal with relatively small,
confined systems and directionality has a strong influence in the
short-time dynamics. 

The initial conditions for the quantum evolution were chosen from a
subset that also minimized finite-size effects. That is, we chose
initial conditions that allowed for the observation of an exponential
decay before the saturation time. For that purpose, we took
$0.5R<x_0<R$, $0.2R<y_0<0.5R$, and initial momentum $\bf{p}_0$ such
that the first collision with the boundary occurred at $x>R$, avoiding
trajectories close to bouncing ball-like modes along $y$. (Such
trajectories were found to lead to strong non-exponential decays in
$M(t)$ for time intervals shorter than the saturation time.)

\begin{figure}
\vspace{.7cm}
\hspace{.25cm}
\epsfxsize=7cm
\epsfbox{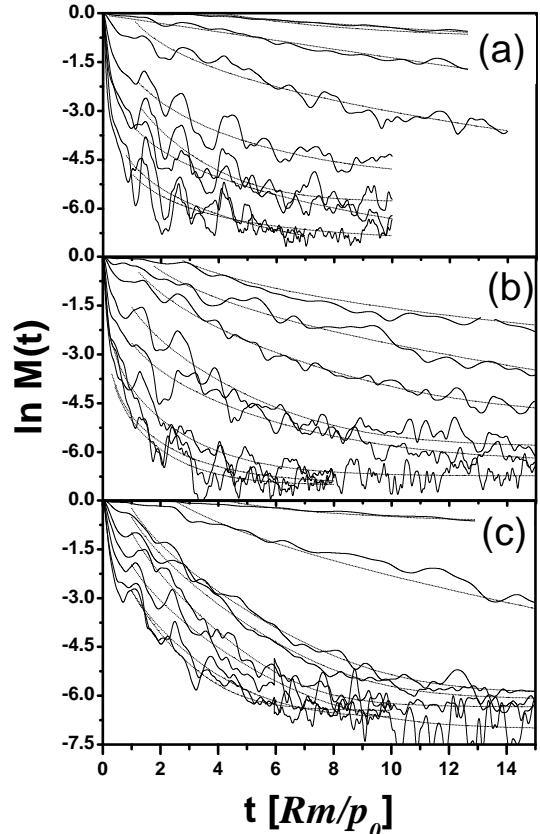}
\caption{$M(t)$ for $\nu=1.5$ (a), $2$ (b), and $3$ (c) for different
values of the perturbation strength: $u = 0.002$, $0.005$, $0.01$,
$0.02$, $0.03$, $0.04$, $0.05$, and $0.06$.}
\label{fig:panels}
\end{figure}

In Fig. \ref{fig:panels} we show $M(t)$ for $\nu=1.5$, $2$, and $3$
for different values of the perturbation strength. In all graphs we
see that the asymptotic decays are approximately exponential within a
certain ranges of $u$, as predicted in
Ref. \onlinecite{Jalabert01}. In order to obtain the characteristic
decay times, we fitted $\ln M(t)$ to the function $\ln [A \exp
(-t/\tau_\phi)/t + M_{\infty}]$. The fit was performed for times $t >
R/v$, where $v = \sqrt{2E/m} = 2$ is the wave packet velocity, to
exclude the initial, non-universal (and non-exponential) time
evolution. It is worth noticing that the usual nonlinear fitting
procedures are rather insensitive to certain combinations of
parameters $\tau_\phi$ and $A$. Thus, while the parameter $M_{\infty}$
could be fixed by averaging the long-time tail of the data, we avoided
the uncertainty in $A$ and $\tau_\phi$ by fixing the value of the
fitted curve at the initial point to be exactly equal to the
respective data value. We checked that such procedure yield values for
$A$ proportional to $u^{-2}$, as expected.

The typical number of samples used in the averaging procedure (for
each trace of the $M(t)$ shown) was in the range 80-100. In fact, we
observed that the number of samples needed to obtain comparable
statistical mean squares fluctuation for $M(t)$ scaled with the
perturbation strength $u$. That is, the larger the perturbation, the
larger the fluctuations in $M(t)$ were. This fact set another
practical limit to the range of perturbation strengths $u$ we could
investigate in our numerical simulations.

\begin{figure}
\epsfxsize=7cm
\vspace{.5cm}
\epsfbox{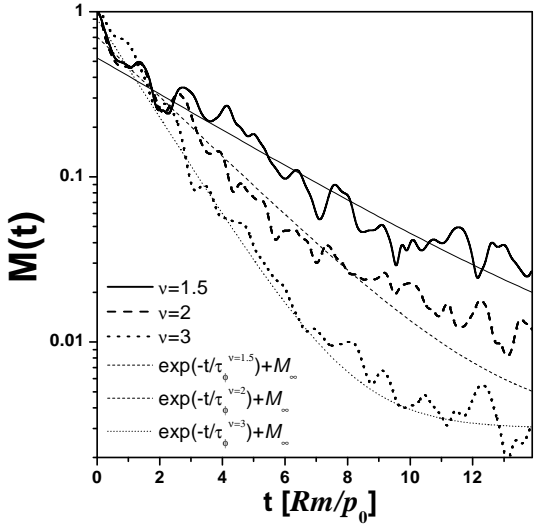}
\vspace{.5cm}
\caption{The fidelity as a function of time. $u=0.01$ for $\nu=1.5$, 
$2$, and $3$. The number of samples used in the averaging behind the 
$\nu=1.5$ curve was 80. 100 samples were used in the two other cases.}
\label{fig:decay}
\end{figure}

In Fig. \ref{fig:decay} we show the fidelity curves for the same
perturbation strength, but different steepness of the confining
potential. Notice that the fluctuations around the (exponential)
fitted curve increase as the billiard walls become softer.


\begin{figure}
\epsfxsize=7cm
\epsfbox{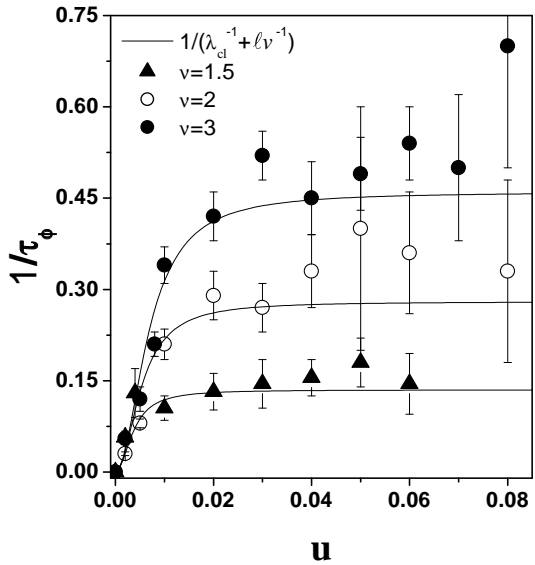}
\vspace{.5cm}
\caption{The characteristic decay rates obtained from
Fig. \ref{fig:panels} as a function of perturbation strength. 
The solid curves correspond to the phenomenological expression,
Eq. (\ref{eq:phenom}).}
\label{fig:decaytimes}
\end{figure}

In Fig. \ref{fig:decaytimes} we have plotted the inverse
characteristic decay times $1/\tau_\phi$ obtained in the fittings as a
function of the impurity strengths $u$ for the three values of
$\nu$. For comparison, we plotted the phenomenological curve
\begin{equation}
\label{eq:phenom}
\tau_{\rm phenom}(u) = \frac{1}{\lambda} + \frac{1}{\alpha_1(u)},
\end{equation}
where $\lambda$ is the classical Lyapunov ($u$ independent) and
$\alpha_1 = v_0/\ell$ is the characteristic decay rate obtained in
Sec. \ref{sec:semiclassicaltheory}. Such curve matches the expected
asymptotic behaviors for $1/\tau_\phi$ at small and large values of
$u$.

The most pronounced feature shared by all data sets is the existence
of a plateau around the classical Lyapunov exponent $\lambda$, as
expected. The semiclassical theory \cite{Jalabert01} predicts that
this saturation should appear when the perturbation is quantum
mechanically strong, but classically weak. This condition, already
presented in Eq. (\ref{constraints}), can be translated into the
inequality $\lambda \ll v/\ell$. Indeed, the results of the
simulations, as presented in Fig. \ref{fig:decaytimes}, are consistent
with the existence of a plateau in $1/\tau_\phi$ for $u$ within this
range. For weak perturbations, the data is also consistent with the
quadratic behavior of $\alpha_1$.

\section{Conclusions}
\label{sec:conclusions}

We studied the time evolution of two wave packets prepared at the same
initial state and time, but evolving under slightly different
Hamiltonians, namely $H_0$ and $H = H_0 + V$. For those systems for
which the Hamiltonian $H_0$ is classically chaotic, the wave packet
overlap decays exponentially in time, according to the semiclassical
theory \cite{Jalabert01}. For the model Hamiltonian introduced in
Sec.\ \ref{sec:model} we numerically verified that the exponential
decay is indeed observed for a broad range of typical strengths of
$V$.

Within the regime of perturbations which are quantum mechanically
strong, but classically weak, the semiclassical theory predicts two
decay laws \cite{Jalabert01}. While the first one is governed by the
mean free path and the wave packet mean velocity, $\alpha_1 =
v_0/\ell$, the second decay law is characterized by the Lyapunov
exponent, namely $\alpha_2 = \lambda(E)$. By estimating the variance
of $V_{nn^\prime}$ we showed that $\alpha_1$ is nothing else than the
Fermi Golden Rule of Ref. \onlinecite{Jacquod01}. Our analytical
findings are in quantitative agreement with the numerical results
obtained from the smooth stadium.

Finally, for sufficiently long times we were able to qualitatively
understand the behavior of the fidelity $M(t)$ as a function of the
strength $u$ of $V$. For very weak $u$, quantum perturbation theory
applies and $\log M(t) \propto -(ut)^2$ \cite{Peres84}. Increasing
$u$, one enters in a regime where although quantum perturbation
theory breaks down, the classical one still holds. We call this the
semiclassical regime. Here, we quantified the crossover from the Fermi
Golden Rule decay to the Lyapunov decay. Finally, by further
increasing $u$, classical perturbation is no longer valid and the
semiclassical approximation ceases to be useful. This picture is
illustrated by Fig. \ref{fig:phasediagram} and nicely numerically
verified by Fig. \ref{fig:decaytimes}. The plateaus obtained in the
simulations show that it is possible to measure the classical Lyapunov
exponent with quantum mechanics over a broad range of perturbation
strengths.

\section{Acknowledgements}

We thank L. Kaplan, T. H. Seligman, and F. Toscano for useful
discussions at the Centro Internacional de Ci\^encias (Mexico). This
work was partially supported by a grant from Funda\c{c}\~ao
VITAE. C.H.L., E.R.M., and R.O.V thank the Brazilian funding agencies
CNPq, FAPERJ, and PRONEX for financial support. H.M.P. is affiliated
to CONICET and F.M.C. is supported by SeCyT-UNC.


\end{multicols}


\begin{thebibliography}{99}
\bibitem{mistake}
  In Ref. \onlinecite{Richter96b}, the result of integral in Eq.\ (3.12),
  taken from $-\infty$ to $\infty$, is missing a factor 2.
\end{thebibliography}
\end{document}